\documentclass[preprint,prd,aps,superscriptaddress,showpacs,showkeys,nofootinbib,notitlepage ]{revtex4-1}
\usepackage{graphicx}
\usepackage{longtable}
\newcommand{\eg}{e.g.,\,}
\newcommand{\ie}{i.e.,\,}
\newcommand{\be}{\begin{equation}}
\newcommand{\ee}{\end{equation}}
\newcommand{\bea}{\begin{eqnarray}}
\newcommand{\eea}{\end{eqnarray}}
\newcommand{\etal}{et al.}

\newcommand{\ra}{\rightarrow}

\newcommand{\pht}{\tilde{\phi}_1}
\newcommand{\chibd}{\frac{\chi_b}{\chi_d}}
\newcommand{\La}{\Lambda r_0^2}
\newcommand{\ch}{\frac{\chi_b}{\chi_d}}

\begin{document}

\bibliographystyle{apsrev}

\title{The Gravitational Lens Equation for Embedded Lenses; Magnification and Ellipticity }

\author{B. Chen}
\email{bin.chen-1@ou.edu}

\affiliation{Homer L.~Dodge Department~of  Physics and Astronomy, University of
Oklahoma, 440 West Brooks,  Norman, OK 73019, USA}
\affiliation{Mathematics Department,University of Oklahoma, 601 Elm Avenue, Norman, OK 73019, USA}

\author{R. Kantowski}
\email{kantowski@nhn.ou.edu}
\affiliation{Homer L.~Dodge Department~of  Physics and Astronomy, University of
Oklahoma, 440 West Brooks,  Norman, OK 73019, USA}

\author{X. Dai}
\email{xdai@ou.edu}
\affiliation{Homer L.~Dodge Department~of  Physics and Astronomy, University of
Oklahoma, 440 West Brooks,   Norman, OK 73019, USA}

\date{\today}

\begin{abstract}
We give the lens equation for light deflections caused by point mass condensations in an otherwise spatially homogeneous and flat universe. We assume the signal from a distant source is deflected by a single condensation before it reaches the observer. We call this deflector an embedded lens because the deflecting mass is part of the mean density. The embedded lens equation differs from the conventional lens equation because the deflector mass is not simply an addition to the cosmic mean.  We  prescribe an iteration scheme to solve this new lens equation and use it to compare our results with standard linear lensing theory.   We also compute analytic expressions for the lowest order corrections  to image amplifications and distortions  caused by incorporating the lensing mass into the mean.

\end{abstract}



\maketitle

\section{Introduction}
%-----------------------------------------------------------

 Conventional extragalactic gravitational lensing assumes that the Universe is homogeneous and isotropic on scales significantly smaller than observer/source/deflector distances, \ie\, that the cosmological principal applies at these distances. It also assumes that a lensing inhomogeneity such as a galaxy or cluster of galaxies is an addition to the homogeneous mean. What we investigate here is the extent to which errors are made because of this latter assumption.  To assume a single galaxy is an addition to the mean might not seem irrational but to assume giant super clusters are is more suspect. In fact they are both contributing to the mean and hence do not act as infinite range deflectors. To understand why, one only has to surround a typical deflector by an imaginary sphere of radius $r$ and note that the average mass density inside the sphere  decreases as $r$ increases until the density reaches the cosmological mean at some $r$=$r_b$.  If this were not correct the cosmological principle would be in error. Beyond the gravitational boundary $r_b$, the gravitational field has returned to the homogeneous mean and the lens ceases to produce any additional deflection of a passing light ray. In this paper we compute modifications to the lens equation caused by this finite range. To make sure we properly account for the lensing  gravity we use an exact solution to Einstein's equations. We assume the deflector is a simple point mass lens embedded in a  flat Friedman-Lema\^itre-Robertson-Walker (FLRW) universe, see Eq.\,(\ref{FLRW}), whose energy content includes pressureless dust (cold dark matter) and a cosmological constant $\Lambda$ ($\Omega_{\rm m}+\Omega_\Lambda=1$). The mathematics of the embedding process is the same as embedding in the
 Swiss cheese cosmological models \cite{Einstein45, Schucking54, Kantowski69, Dyer74}. These models are the only known exact general relativistic (GR) solutions which embed spherical inhomogeneities into homogeneous background universes. The range $r_b$ above is given by the comoving radial boundary of the homogeneous sphere that has been replaced by the condensation. Beyond that boundary the gravity caused by a condensation and a homogeneous sphere are exactly the same. Sch\"ucker  \cite{Schucker09a} refers to this radius as the Sch\"ucking radius. For a point mass lens the removed dust sphere of comoving radius $\chi_b$ is replace by a Kottler condensation \cite{Kottler18}, \ie Schwarzschild with a cosmological constant, see Eq.\,(\ref{Kottler}). In \cite{Kantowski10, Chen10} we derived analytical expressions for the bending angle $\alpha$ and the time delay $\Delta T$ of a photon that encounters such a condensation. Related work appeared in \cite{Rindler07,Schucker09b,Schucker10,Boudjemaa,Ishak10a,Ishak10b}. In this paper we derive the embedded lens equation and prescribe a scheme to iteratively solve it.

The flat FLRW metric for the background cosmology can be written as
\be
ds^2=-c^2dT^2+R(T)^2\left[{d\chi^2}+\chi^2(d\theta^2+\sin^2\theta d\phi^2)\right],
\label{FLRW}
\ee
and the embedded condensation is described by the Kottler or Schwarzschild-de Sitter metric \cite{Kottler18} which can be written as
 \be
 \label{Kottler}
ds^2=-\gamma(r)^{-2}c^2dt^2+
\gamma(r)^2dr^2+
r^2(d\theta^2+\sin^2\theta\, d\phi^2),
\ee
where $\gamma^{-1}(r)\equiv\sqrt{1-\beta^2(r)}$ and $\beta^2(r)\equiv r_s/r+\Lambda r^2/3$.
The constants  $r_s$ and $\Lambda$  are the Schwarzschild radius  ($2G{\rm m}/c^2$) of the condensed mass and the cosmological constant respectively.
By matching the first fundamental forms  at the Kottler-FLRW boundary, angles $(\theta,\phi)$ of Eqs.\,(\ref{Kottler}) and (\ref{FLRW}) are identified and the expanding Kottler radius $r_b$ of the void is related to the comoving FLRW radius $\chi_b$ by
\be\label{rb-T}
r_b=R(T)\chi_b.
\ee
By matching the second fundamental forms  the Schwarzschild radius $r_s$ of the Kottler condensation is related to FLRW by
\be\label{second-form}
r_s=\Omega_{\rm m}\frac{H_0^2}{c^2}(R_0\chi_b)^3,
\ee
where  $H_0$ is the Hubble constant and the cosmological constant $\Lambda$ is constrained to be the same inside and outside of the Kottler hole.

In Section II we give the lens equation valid for deflections caused by Kottler condensations in the flat FLRW universe and numerically compare its predictions with conventional lensing theory for a source at redshift one
 and a deflector at redshift one half. In Section III we give analytic expressions for image magnifications and distortions for the embedded point mass lens (to lowest order only) and compare them with conventional lensing results.

\section{ The Lens Equation}

\begin{figure*}
\includegraphics[angle=90,width=0.7\textwidth,height=0.28\textheight]{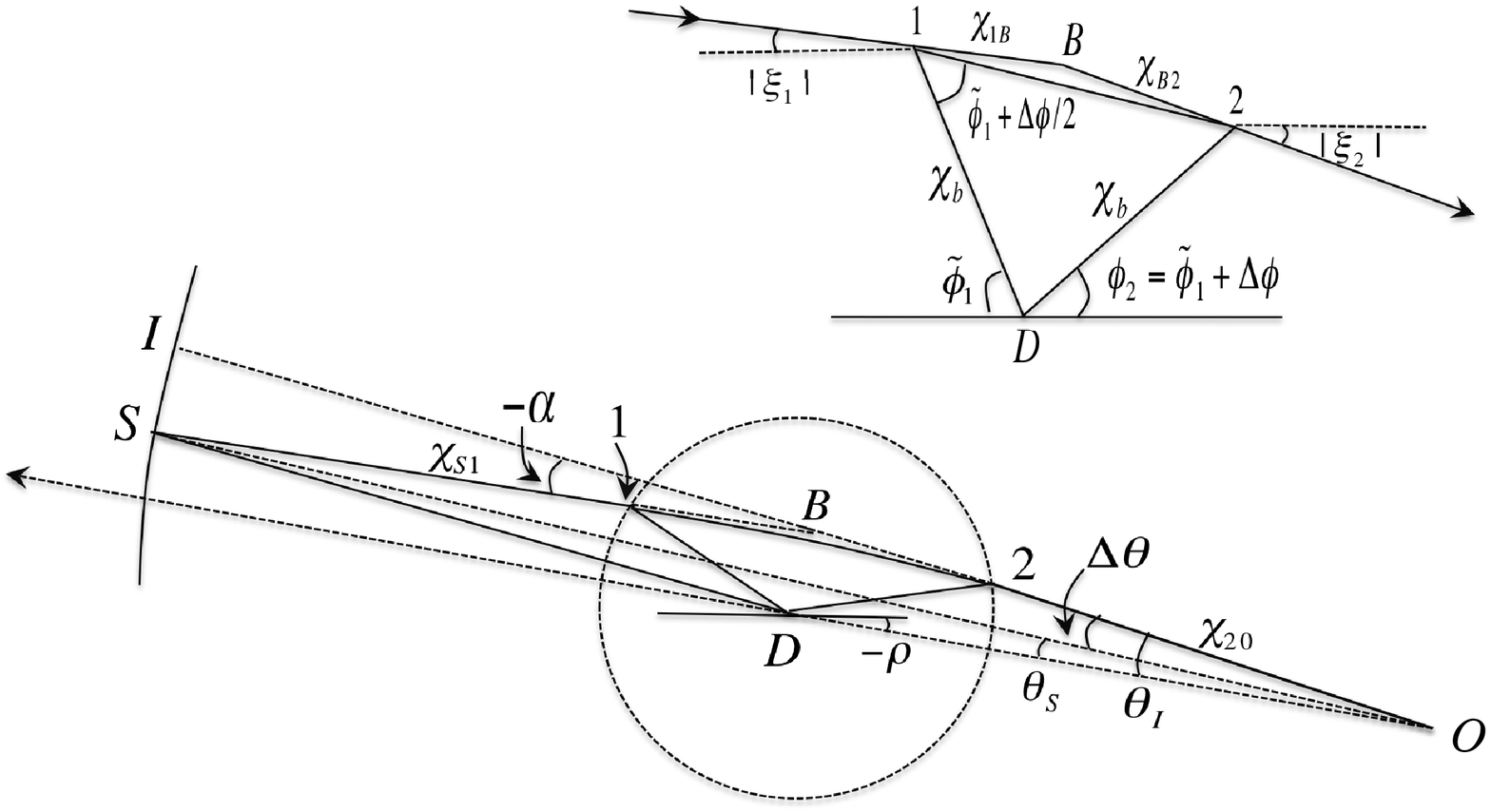}
\caption{The  comoving embedded lensing geometry. Points $S, D$ and $O$ represent respectively the source, deflector, and observer positions.  The point $B$ is a fictitious reflection point. Points $1$ and $2$ denote the photon's entrance and exit from the Kottler void. The bending angle is $\alpha$, $\theta_I$ and $\theta_S$ are respectively the image and source position angles at the observer measured relative to the optical axis $OD$, and $\Delta\theta\equiv \theta_I-\theta_S.$ A similar geometry appears in  Fig.\,1 of \cite{Kantowski10}.  The figure represents the $\theta =\pi/2$  plane (the plane containing the photon's orbit) of the spherical polar coordinates used in Eqs.\,(\ref{FLRW}) and (\ref{Kottler}). The $\phi$ orientation is fixed by requiring the photon's point of closest approach to the Schwarzschild mass, $r_0$, occur at $\phi = \pi/2$.}
\label{fig:lens}
\end{figure*}

The Swiss cheese lensing geometry is shown in Fig. \ref{fig:lens}. The deflected photon leaves a source $S$, enters a Kottler hole at point 1, exits at point 2 with a deflection angle $\alpha<0$, and then proceeds to the observer at 0.
Point $B$ is the intersection of the forward and backward extensions of respective FLRW rays $S1$ and ${20}$ drawn as if the  Kottler hole were absent and the original ray was simply reflected at point $B$. Angles $\theta_I$ and $\theta_S$ are respectively the image and source positions relative to the observer-deflector optical axis  $OD$. The rotation angle $\rho$ measures the difference between the horizontal axis [with respect to which we measure the spherical polar angle $\phi$, see Eqs.\,(\ref{Kottler}) and (\ref{rho}) and Fig. 1]  and the optical axis. A negative $\rho$ is a clockwise rotation of the observer. The lens equation for a given deflector mass and background cosmology is simply the equation that gives $\theta_I$ as a function of  $\theta_S$ for fixed comoving source-observer distance $\chi_s$ and deflector-observer distance $\chi_d$, and fixed photon  arrival time $T_0$. For non-embedded lenses, \ie for conventional linear lensing theory, this relation is straightforward to obtain even for complicated lensing mass profiles, because the deflector is completely unrelated to the cosmology. For an embedded lens this is no longer the case. However, because of the azimuthal symmetry of the lensing geometry all photon orbit variables can be thought of as depending on a single independent variable.  Choosing $\theta_S$ or the photon's minimum Kottler coordinate $r_0$ would be logical but not convenient. In what follows we have chosen to give all quantities as functions of $\pht$ where  $\pi-\pht$ is  the azimuthal angle of the photon at entry into the Kottler void (see Fig.\,1, or Fig.\,1 of \cite{Kantowski10}). Because $r_0(\pht)$ is a complicated function, $r_0$ is retained in all expressions and only evaluated when needed.

The embedded lens equation can be obtained by applying the law of sines to the triangle SB0 of Fig.\,1
\be\label{lens-eq-1}
\sin(\theta_S-\theta_I-\alpha)=\frac{\chi_{B0}}{\chi_s}\sin(-\alpha).
\ee
 The comoving distances $\chi_s$ and $\chi_d$ are often replaced by angular diameter distances $D_s$ and $D_d$ which are respectively functions of redshifts $z_s$ and $z_d$.
The embedded lens equation can be compared to the standard linear lensing equation \cite{Bourassa75,Ehlers} for flat FLRW
\be\label{linear-lens}
 \theta_S-\theta_I=-\frac{D_{ds}}{D_s}(-\alpha)=-\frac{\chi_s-\chi_d}{\chi_s}(-\alpha),
 \ee
where small angle approximations are made and the differences between distances from the observer to the deflector and to the reflection point $B$ ($\chi_d$ and  $\chi_{B0}$) are neglected. Since we are now computing the linear and non-linear corrections to the standard lensing theory, we cannot make such simplifications as is done in \cite{Ishak08a} and \cite{Sereno09}. To find the relation between these two distances we use the
comoving triangle $D20$ and obtain
\bea
\chi_{20}&=&\left[\cos\theta_I-\cos(\pht-\xi_1+\Delta\phi-\alpha)\chibd\right]\chi_d,
\label{chi20}
\eea
where  $\chi_b$, the Kottler void radius [see Eq.\,(\ref{second-form})], is assumed known. The angles   $\xi_1$, $\xi_2=\xi_1+\alpha,$ $\pht,$ and $\Delta \phi$ are exhibited in Fig. 1 and are the same as those used in \cite{Kantowski10, Chen10} where analytic expansions for them as explicit functions of $r_0$ and $\pht$ can  be found. The angles $\xi_1$ and $\xi_2$ are negative and give the respective slopes of the photon as it enters the Kottler hole at azimuthal angle $\pi-\pht$  and exits at angle $\phi_2=\Delta\phi+\pht$ (see Fig.\,1 of \cite{Kantowski10}).
The comoving distance $\chi_{B2}$ can be obtained from trig identities applied to triangles $1B2$ and $1D2$ of Fig.\,1
\be\label{locate-D}
\chi_{B2}=-2\frac{\sin(-\Delta\phi/2+\xi_1)}{\sin\alpha}\cos\left(\pht+\frac{\Delta\phi}{2}\right)\chi_b.
\ee
Combining this with Eq.\,(\ref{chi20}) we obtain the relation of $\chi_{B0}$ to $\chi_d$,
\bea
\chi_{B0}&\equiv&\chi_{20}+\chi_{B2}\cr
&=&\Bigg\{\cos\theta_I-\Bigg[\cos(\pht-\xi_1+\Delta\phi-\alpha)\cr
& &+\frac{2\sin(-\Delta\phi/2+\xi_1)}{\sin\alpha}\cos\left(\pht+\frac{\Delta\phi}{2}\right)\Bigg]\chibd\Bigg\}\chi_{d}\cr
&\equiv&g(\theta_I, \xi_1,\Delta\phi, \alpha)\chi_d.
\label{g}
\eea
The new lens equation (\ref{lens-eq-1}) becomes
\be\label{lens-eq-2}
\theta_S=\theta_I+\alpha+\sin^{-1}\left[\frac{\chi_d}{\chi_s}\ g(\theta_I, \xi_1,\Delta\phi, \alpha)\sin(-\alpha)\right],
\ee
The task at hand is to evaluate all variables on the right hand side of Eq.\,(\ref{lens-eq-2}) as functions of a common variable \eg \,$\pht$. Once accomplished, $\theta_S(\pht)$ and  $\theta_I(\pht)$ can be tabulated to give the desired image position as a function of the source position, $\theta_I(\theta_S)$.
The image angle  $\theta_I$ can be determined from knowledge of $\xi_1$, $\Delta\phi,$ and $\alpha$ by applying the law of sines to the triangle D20
\be\label{theta-I}
\sin\theta_I=\sin(\pht-\xi_1+\Delta\phi-\alpha)\frac{\chi_b}{\chi_d}.
\ee
The bending angle $\alpha$ is  given by Eq.\,(32) of \cite{Kantowski10},  $\Delta\phi\equiv \phi_2-\pht$ is given by Eq.\,(13) of \cite{Chen10}, and the photon's slope angle $\xi_1$ results from evaluating Eqs.\,(16)-(19) of \cite{Kantowski10} at the photon's entry point into the Kottler void (to fourth order)
\bea
\xi_1&=&-\beta_1\sin\pht+\frac{m}{r_0}\cos\pht(2+\sin^2\pht)-\frac{1}{3}\beta_1\frac{m}{r_0}(6\cr
&&-3\sin^2\pht-2\sin^4\pht)-\frac{1}{18}\beta_1\Lambda r_0^2\sin\pht -\frac{1}{4}\frac{m^2}{r_0^2}\cr
& &\times\Big[15(\pht-\frac{\pi}{2})+\cos\pht\big(8-15\sin\pht+4\sin^2\pht\cr
& &+14\sin^3\pht+4\sin^5\pht\big)\Big]+{\cal O}(5).
\label{xi_1}
\eea
The rotation angle $\rho$ can be computed from the photon's exiting slope $\xi_1+\alpha$ and the image position $\theta_I$ using
\be
\rho=\xi_1+\alpha+\theta_I \hspace{4pt}  <0.
\label{rho}
\ee

The expansion speed $\beta=v/c$  of the void boundary relative to stationary Kottler observers  is defined in Eq.\,(\ref{Kottler}) and when evaluated at the photon's entry point is called $\beta_1$ (see Fig.\,1 of \cite{Kantowski10}). Keeping terms to $4^{\rm th}$ order is necessary in order to correct point mass time delays for embedding.

In the expressions for $\xi_1,\Delta\phi,$ and $\alpha$, approximation orders have been counted as follows: $\beta_1$ is $1^{\rm st}$ order, $r_s/r_0$ and $\Lambda r_0^2$ are both $2^{\rm nd}$. All terms are made of sums and/or products of these. The expansion speed $\beta_1$ depends on $\pht$ and $r_0$ through its dependence on $r_1$  (which is given by the symmetric null geodesics of the Kottler metric  Eq.\,(\ref{Kottler}))
\bea\label{r-phi}
r_1&&=\frac{r_0}{\sin\pht}\Bigg\{1+\frac{r_s}{2 r_0}\left(1+\sin\pht-\frac{2}{\sin\pht}\right)-\left(\frac{r_s}{2 r_0}\right)^2\times\cr
&&\left[\frac{17}{4}-\frac{1}{4}\sin^2\pht-\frac{4}{\sin^2\pht}+\frac{15}{8}\left(\pi-2\pht\right)\cot\pht\right]\cr
&&+{\cal O}(6)\Bigg\}.
\eea
The above expansion is valid only when $\sin\pht \gg r_s/r_0$.
 All quantities on the RHS of the embedded lens equation (\ref{lens-eq-2}) can now be evaluated as functions of $\pht$ and $r_0$. These two variables fix the photon's symmetric orbit (symmetric about $\phi=\pi/2$) while in the Kottler hole. They are independent unless the photon is additionally constrained by originating at a specific cosmic source or arriving at a specific observer. To eliminate one of these two variables an additional relation between them  such as a  cosmic timing constraint must be used.  For the photon which started at a fixed $\chi_s$ to have reached the observer at time $T_0$ after entering the Kottler void at $\pht$ and passing with minimum impact $r_0$,  it must have impacted the Kottler void at a specific time $T_1$ or equivalently at a specific redshift $z_1$ ($1+z_1= R_0/R(T_1)$).
Knowledge of $z_1$ allows us to independently determine $r_1$ from the embedding equations (\ref{rb-T}) and (\ref{second-form}) \ie by using
\be\label{rb-z}
r_1=\frac{1}{1+z_1}\left(\frac{r_s}{\Omega_{\rm m}}\frac{c^2}{H_0^2}\right)^{1/3}.
\ee
Because  $z_1$ is not assumed known we compute $z_1-z_d$, the difference in entry redshift and the (assumed known) deflector redshift, using techniques similar to those developed in \cite{Kantowski10, Chen10}.  The result  up to fourth order is
\bea\label{zdz1}
z_d-z_1&=&(1+z_1)\Big[\Delta z^{\rm 1 st}(z_1,\pht)+\Delta z^{\rm 2 nd}(z_1,r_0,\pht)\cr
&&+\Delta z^{\rm 3rd}(z_1,r_0,\pht)+\Delta z^{\rm 4 th}(z_1,r_0,\pht)\Big],
\eea where
\be\label{first}
\Delta z^{\rm 1 st}=-\beta_1 \cos\pht ,
\ee
\bea\label{second}
\Delta z^{\rm 2 nd}&=& -\frac{\La}{3}+\frac{1}{2}\beta_1\ch\sin^2\pht+\frac{1}{2}\frac{m}{r_0}\sin\pht\Big(3\cr
&&-7\sin^2\pht\Big),
\eea
 \bea
\Delta z^{\rm 3 rd}&=&\frac{1}{6}\beta_1\La\cos\pht-\frac{\La}{3}\ch\cos\pht+\frac{1}{3}\beta_1\frac{m}{r_0}\cr
&\times&\Bigg(\cos\pht\left[7+26\sin\pht^2\right]+12\log\tan\frac{\pht}{2}\Bigg)\sin\pht\cr
 &-&\frac{7}{2}\frac{m}{r_0}\ch\cos\pht\sin^3\pht,
 \eea
and
\bea
\Delta z^{\rm 4 th}&=&\frac{1}{6}\beta_1\La\ch(1-2\sin\pht^2)
+\frac{1}{2}\beta_1\ch\frac{m}{r_0}\Big(4+9\sin\pht^2\cr
 &-&18\sin\pht^4\Big)\sin\pht
 +\frac{1}{8}\beta_1\left(\ch\right)^3\sin^4\pht+ \frac{3}{8}\frac{m}{r_0}\cr
 &\times&\left(\ch\right)^2\sin^5\pht
 -\frac{1}{36}\frac{m}{r_0}\La \csc\pht
 \Big(61+24\sin\pht\cr
 &+&124\sin\pht^2-227\sin\pht^4 +48\cos\pht\log\tan\frac{\pht}{2}\Big)\cr
 &+&\frac{1}{12}\frac{m^2}{r_0^2}
 \Big(36-18\sin\pht-431\sin\pht^2\cr
 &+&42\sin\pht^3-188\sin\pht^4+595\sin\pht^6\cr
 &-&240\cos\pht\sin^2\pht\log\tan\frac{\pht}{2}\Big).
\eea
In the above
\be
\frac{\chi_b}{\chi_d}=\frac{1}{1+z_d}\left(\frac{r_s}{\Omega_{\rm m}}\frac{c^2}{H_0^2}\right)^{1/3}\frac{1}{D_d},
\ee
is taken as an additional small parameter no larger than $1^{\rm st}$ order.

Equations (\ref{zdz1}), (\ref{r-phi}) and (\ref{rb-z})  are three equations relating four variables $z_1,$ $r_0$,  $r_1$, and  $\pht$. They can be solve iteratively (four iterations) giving  $z_1,$ $r_0$, and $r_1$ as functions of $\pht$. For an example, to obtain $z_1$ correct to the first order in $\beta_1,$ we use Eqs.\,(\ref{zdz1}) and (\ref{first})
\be
z_1=z_d-(1+z_d)\Delta z^{\rm 1st}(z_d,\pht)=z_d+(1+z_d)\beta(z_d)\cos\pht,
\ee this can be inserted into Eq.\,(\ref{rb-z}) to obtain $r_1$ correct to first order in $\beta_1.$ This $r_1$ is then inserted into Eq.\,(\ref{r-phi}) (only the lowest order is needed here, \ie $r_0=r_1\sin\pht$) to obtain $r_0$ correct to first order. For the next iteration, we include Eq.\,(\ref{second}) and the $r_s/r_0$ term in Eq.\,(\ref{r-phi}), and so on. With $z_1(z_d,\pht),$ $r_0(z_d,\pht)$ and $r_1(z_d,\pht)$ in hand, we can compute $\theta_I,$ $\xi_1,$ $\Delta\phi$, and $\alpha$ in terms of $\pht$ and finally solve the embedded gravitational lensing equation (\ref{lens-eq-2}) for $\theta_S(\pht)$ which can be tabulated to give $\theta_S(\theta_I)$ for a given image.

\begin{figure*}
\begin{center}$
\begin{array}{cc}
\includegraphics[width=0.4\textwidth,height=0.25\textheight]{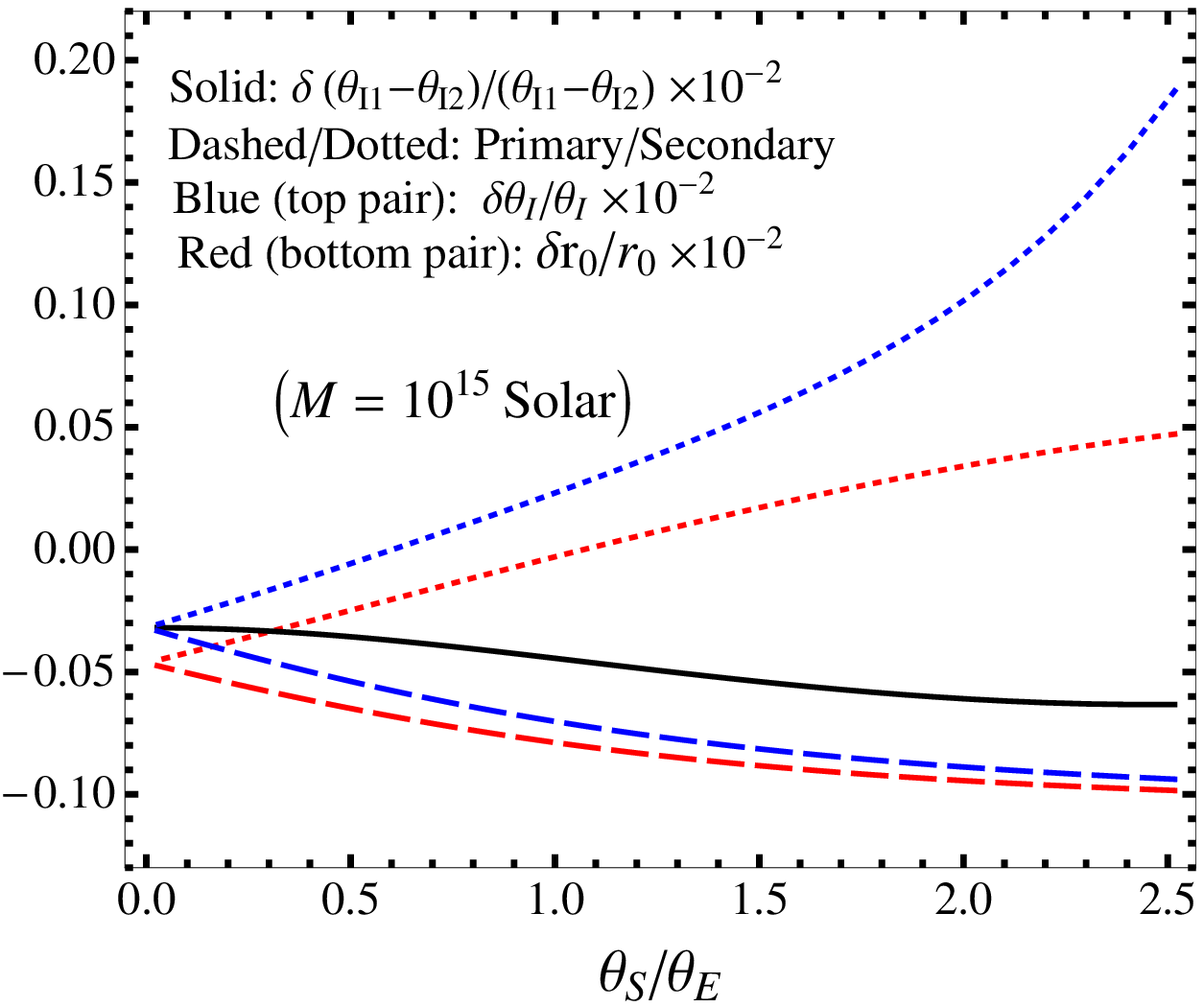}
\hspace{10pt}
\includegraphics[width=0.4\textwidth,height=0.25\textheight]{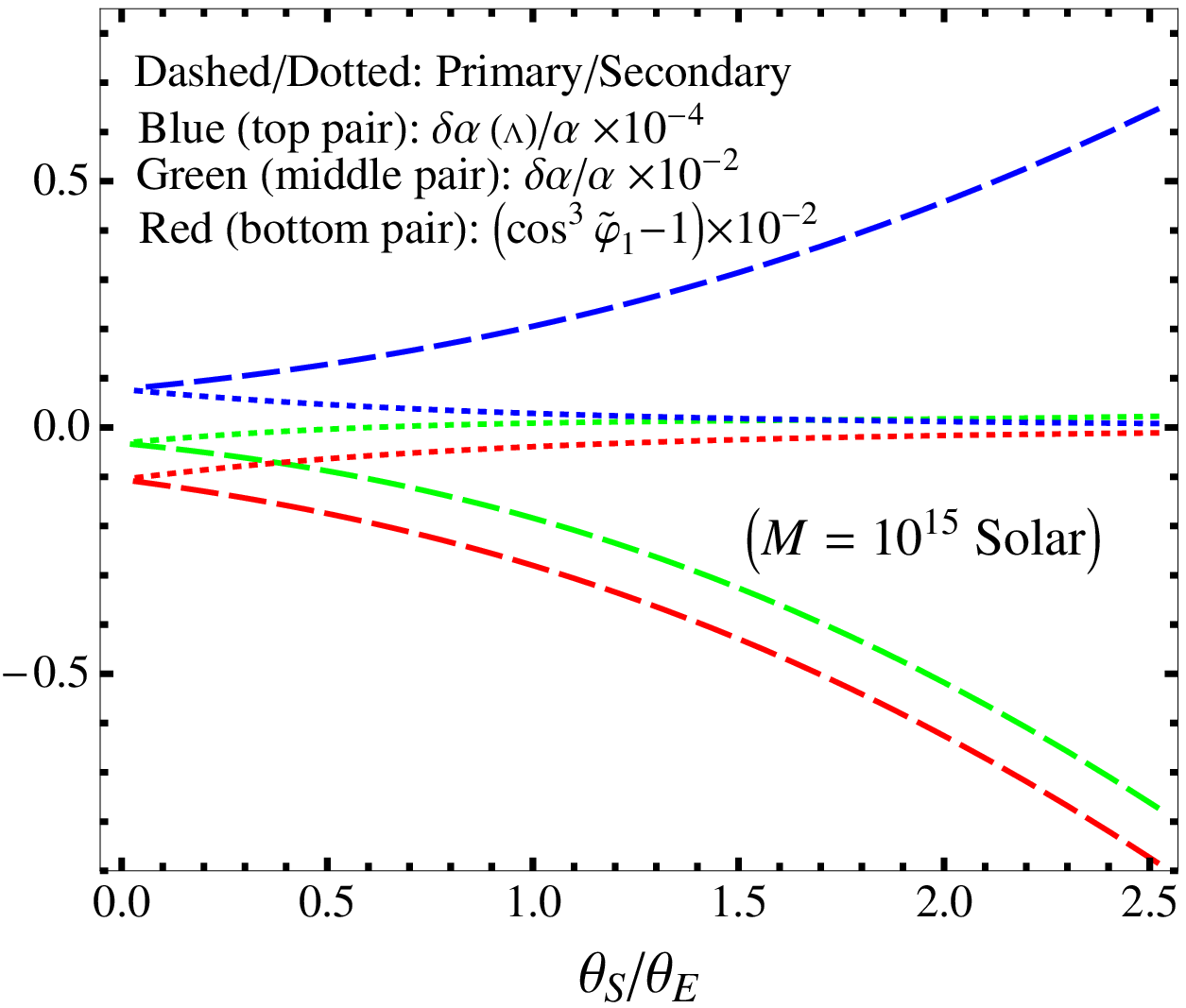}
\end{array}$
\end{center}
\caption{The embedded point mass lens versus the Schwarzschild lens.  The deflector/source redshifts are respectively $z_d=0.5,$ $z_s=1.0$; the cosmological parameters are $\Omega_{\rm m}=0.3,$  $\Omega_\Lambda=0.7,$  and $H_0=70\,{\rm km}\,{\rm s}^{-1}\,{\rm Mpc}^{-1};$ and the deflector mass is ${\rm m}=10^{15} \>M_{\odot}.$   The abscissa $\theta_S$ is  the source angle measured in units of the classical Einstein ring angle $\theta_E$ and  the dashed/dotted lines are for primary/secondary images. The bifurcating blue curves are above the corresponding  bifurcating red curves. The green bifurcating pair of curves in the right panel are between the upper blue pair and lower red pair. The solid curve in the left panel measures the relative correction of the angle between the primary and secondary images.    }
\label{fig:M15}
\end{figure*}

\begin{figure*}
\begin{center}$
\begin{array}{cc}
\includegraphics[width=0.4\textwidth,height=0.24\textheight]{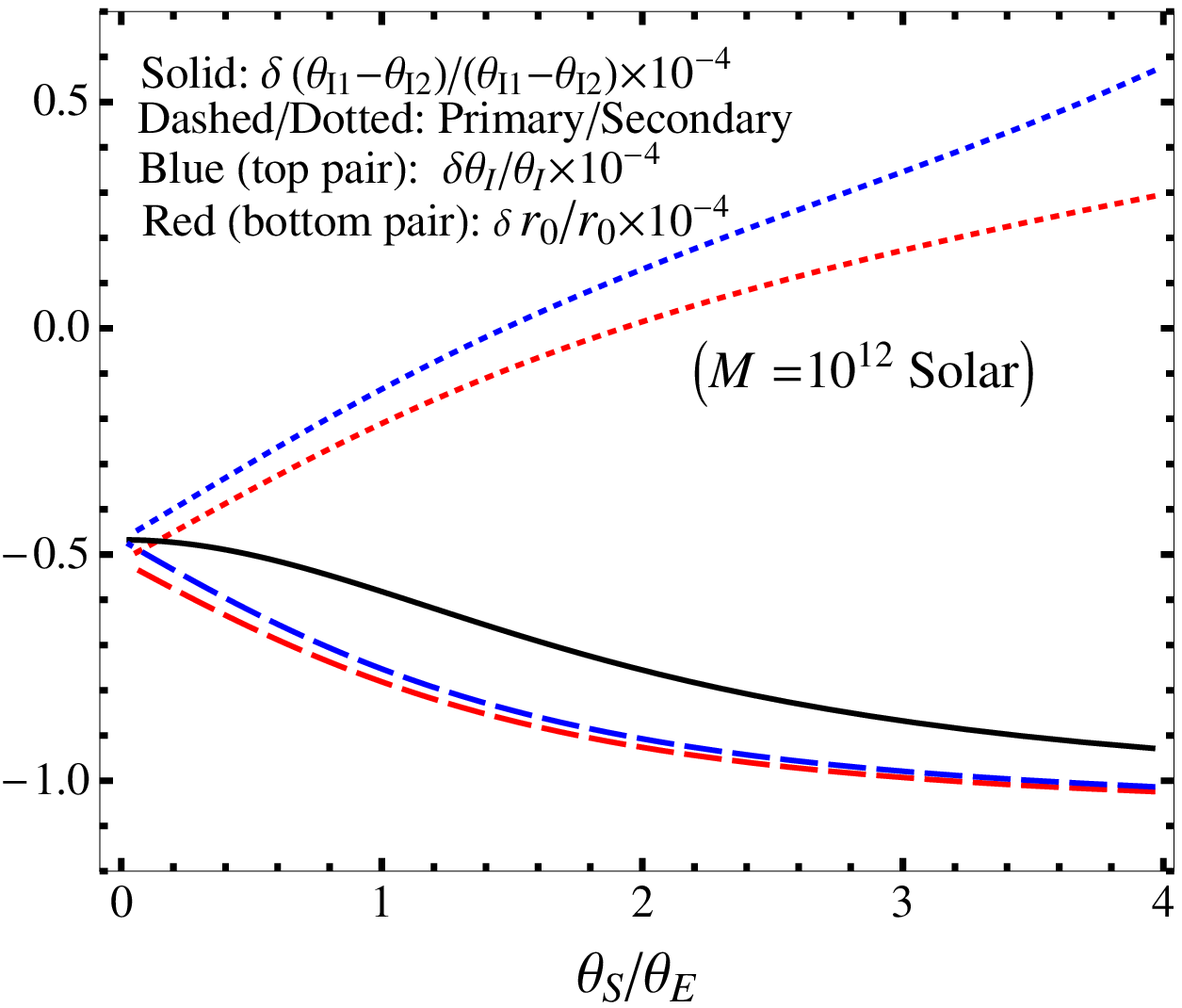}
\hspace{10pt}
\includegraphics[width=0.4\textwidth,height=0.24\textheight]{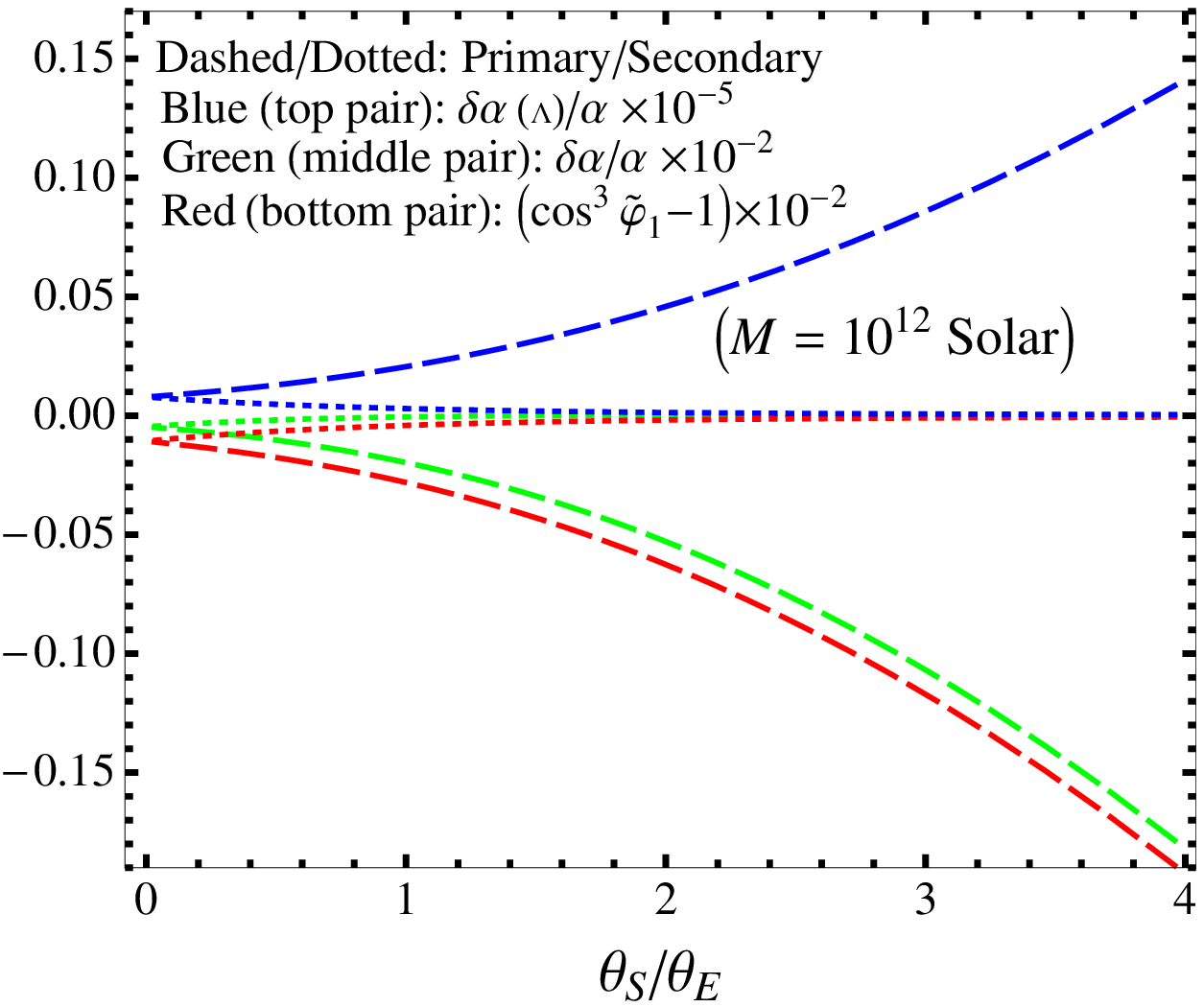}
\end{array}$
\end{center}
\caption{The embedded point mass lens versus the Schwarzschild lens. Same as Fig.\,\ref{fig:M15}, except that the deflector mass ${\rm m}=10^{12} \>M_{\odot}.$ }
\label{fig:M12}
\end{figure*}

In Figs.\,\ref{fig:M15} and \ref{fig:M12} we have solved the embedded point mass Swiss cheese lens equation (\ref{lens-eq-2}) and compared the results with those of  the conventional Schwarzschild point mass lensing theory.   We chose deflector/source redshift  respectively $z_d=0.5,$ $z_s=1.0,$ cosmological parameters  $\Omega_{\rm m}=0.3,$  $\Omega_\Lambda=0.7,$  and $H_0=70\,{\rm km}\,{\rm s}^{-1}\,{\rm Mpc}^{-1}.$   In Fig.\,\ref{fig:M15},  we chose a deflector mass ${\rm m}=10^{15} \>M_{\odot}$ (a rich cluster). For each source angle $\theta_S,$ we solved Eq.\,(\ref{lens-eq-2}) using the iteration scheme described above obtaining  $\pht,$ $z_1,$ $r_0$, $r_1,$ $\theta_I,$ etc., for both the primary and secondary images.   The conventional Schwarzschild results are given by Eq.\,(\ref{linear-lens}). The impact parameter in conventional lensing is simply taken as $r_{0({\rm Sch})}=\theta_{I({\rm Sch})}D_{d}.$
The dashed/dotted  curves are for primary/secondary images, and the solid curve is the correction to the angle between  image pairs, \ie $\theta_{I1}-\theta_{I2}.$
In the left panel, we  compute
 the relative correction in the image position, \ie $\delta \theta_{I}/\theta_{I({\rm Sch})}$ (blue-upper bifurcating pair of curves),
 and
the relative correction of the impact parameter $r_0, $
\ie $\delta r_0/r_{0({\rm Sch})}$,
 where $\delta r_0\equiv r_{0}-r_{0({\rm Sch})}$ (red-lower bifurcating pair of curves).

 In the right panel, we compute
 the net correction of the bending angle $\alpha$ (central pair of green curves),
 the effect of the linear correction alone, \ie $\cos^3\pht-1$ (lower pair of red curves),
 and
 the contribution of the cosmological constant $\Lambda$ (upper pair of blue curves).
 Figure \ref{fig:M12} is the same as Fig.\,\ref{fig:M15} except that it is for ${\rm m}=10^{12} \>M_{\odot}$ (a typical large galaxy).  For ${\rm m}=10^{15} \>M_{\odot}$,  corrections in the image angle $\theta_I$ can be as large as $0.2\%,$ and corrections in the bending angle $\alpha$ can be as large as $-0.8\%.$ For ${\rm m}=10^{12} \>M_{\odot}$, corrections in the image angle $\theta_I$ can be as large as $0.01\%,$ and corrections in the bending angle $\alpha$ can be as large as $-0.18\%.$
\section{Image Magnification and Ellipticity}

In this section we include only the lowest order correction to the standard lensing equation caused by the finite range of the embedded point mass Swiss cheese lens. Sereno \cite{Sereno08} computes  alterations in the magnification but only within the Kottler void. We assume $\sin\theta_I\ll 1,$ $\sin\theta_S\ll 1,$ and that the Kottler hole is much smaller than the observer-deflector distance, \ie\,$g(\pht)- 1\ll 1$, see Eq.\,(\ref{g}). From Eq.\,(\ref{lens-eq-1}) we obtain
\be
\theta_S-\theta_I=-\frac{D_{ds}}{D_s}(-\alpha),
\ee which is the same as the standard lens equation (\ref{linear-lens}) except that the bending angle to the lowest order now contains a $\cos^3\pht$ factor caused by the finite range of the deflector
\be
\alpha=-2\frac{r_s}{r_0}\cos^3\pht,
\ee
see Eq.\,(32) of \cite{Kantowski10}. Equation (\ref{lens-eq-1}) is the form assumed correct by \cite{Sereno09} but with a different expression for the deflection angle $\alpha$.

To lowest order the minimum Kottler impact is
\be
r_0 = D_{d}\theta_I+{\cal{O}}(\beta_1),
\ee
[see Eqs.\,(\ref{rb-T}), (\ref{theta-I}) and (\ref{r-phi})] and the embedded lens equation to lowest order becomes
\be\label{lowest-appro}
\theta_S-\theta_I=-\frac{\theta^2_E}{\theta_I}\cos^3\pht.
\ee
The angle $\theta_E$ is the familiar Einstein ring radius
\be
\theta_{\rm E}\equiv \sqrt{2\frac{D_{ds}r_s}{D_dD_s}},
\ee and  from Eq.\,(\ref{theta-I}) $\theta_I$ is related to $\pht$  by 
\be\label{phi-thetaI}
\sin\pht=\frac{\theta_I}{\chi_b/\chi_d}+{\cal O}(\beta_1).
\ee
This gives us a modified  Einstein ring radius (to lowest order)
\be
\theta'_{\rm E}=\sqrt{2\frac{D_{ds}r_s}{D_dD_s}}(\cos\pht)^{3/2},
\ee
(see \cite{Ishak08b} for modifications in the Einstein ring within the Kottler void).
The two images for the standard point mass lens are easily found at
\be
\theta_I^\pm=\frac{1}{2}\left\{\theta_s\pm\sqrt{\theta_s^2+4\theta_E^2}\right\},
\label{Schwarzschild-images}
\ee
however, to find the corresponding image positions for the embedded lens you must solve Eq.\,(\ref{Schwarzschild-images}) with $\theta_E$ replaced by $\theta'_{\rm E}$.

The amplification and shear for the embedded lens can be found by a familiar \cite{Ehlers} rescaling ($\theta_S\ra \theta_S/\theta_E\equiv y,$ $\theta_I\ra \theta_I/\theta_E\equiv x$).
Equation (\ref{lowest-appro}) simplifies to
\be
{\bf y}={\bf x}-\frac{\cos^3\pht}{x^2}{\bf x},
\ee where
\be
\sin\pht=\frac{x}{(\chi_b/\chi_d)/\theta_E}.
\ee
The 2-d Jacobian $A\equiv\partial{\bf y}/\partial {\bf x}$ is found to be \cite{Bourassa75}
\bea
A&=&\left(1-\frac{\cos^3\pht}{x^2}\right)\left[\matrix{1 & 0 \cr 0  & 1 }\right]\nonumber\\
&&+\frac{\cos\pht(2+\sin^2\pht)}{x^4}\left[\matrix{x_1^2 & x_1x_2 \cr x_1x_2  &  x_2^2 }\right],
\eea
which has two eigenvalues
\bea
a_1&=&1+\cos\pht(1+2\sin^2\pht)\frac{1}{x^2}, \nonumber \\
 a_{2}&=&1-\cos^3\pht \frac{1}{x^2}.
\eea
Writing
\be
A=\left(\matrix{1-\kappa-\gamma_1 & -\gamma_2 \cr -\gamma_2  &  1-\kappa+\gamma_1}\right)
\ee
as is commonly done in standard gravitational lensing theory, we immediately obtain a negative surface mass density
\be
\kappa = -\frac{3}{2}\sin^2\pht\cos\pht \frac{1}{x^2},
\ee
and two shear components
\bea\label{shear}
\gamma_1&=&-\cos\pht(2+\sin^2\pht)\frac{x_1^2-x_2^2}{2x^4},\cr
\gamma_2&=&-\cos\pht(2+\sin^2\pht)\frac{x_1x_2}{x^4},
\eea with total shear
\be
\gamma\equiv \sqrt{\gamma_1^2+\gamma_2^2}=\cos\pht(2+\sin^2\pht)\frac{1}{2x^2}.
\ee
The amplification  $\mu$ for an image is given by
\bea\label{mu}
\mu^{-1}({\bf x})&=&\det A =(1-\kappa)^2-\gamma^2=a_1a_2\cr
&=& 1+3\cos\pht\sin^2\pht\frac{1}{x^2}-\cos^4\pht(1+2\sin^2\pht)\frac{1}{x^4}.\nonumber
\eea The image of a circular source (eccentricity $\epsilon=0$) will be an ellipse of eccentricity
\be
\epsilon=\sqrt{1-\frac{a_2^2}{a_1^2}}=\frac{\sqrt{(2x^2+3\sin^2\pht\cos\pht)(2+\sin^2\pht)\cos\pht}}{x^2+\cos\pht(1+2\sin^2\pht)}.
\ee

\begin{figure*}
\begin{center}$
\begin{array}{cc}
\includegraphics[width=0.4\textwidth,height=0.24\textheight]{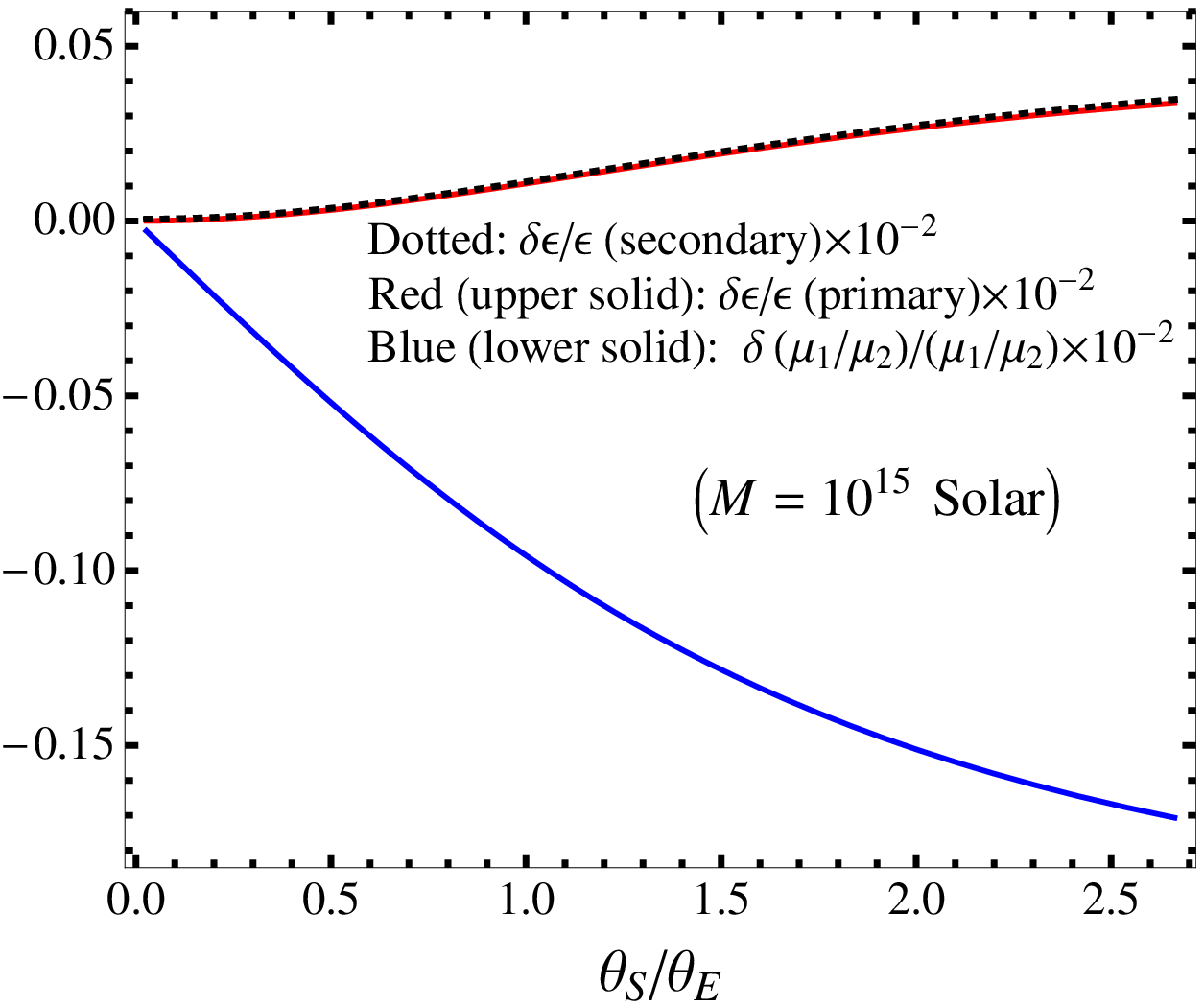}
\hspace{10pt}
\includegraphics[width=0.4\textwidth,height=0.24\textheight]{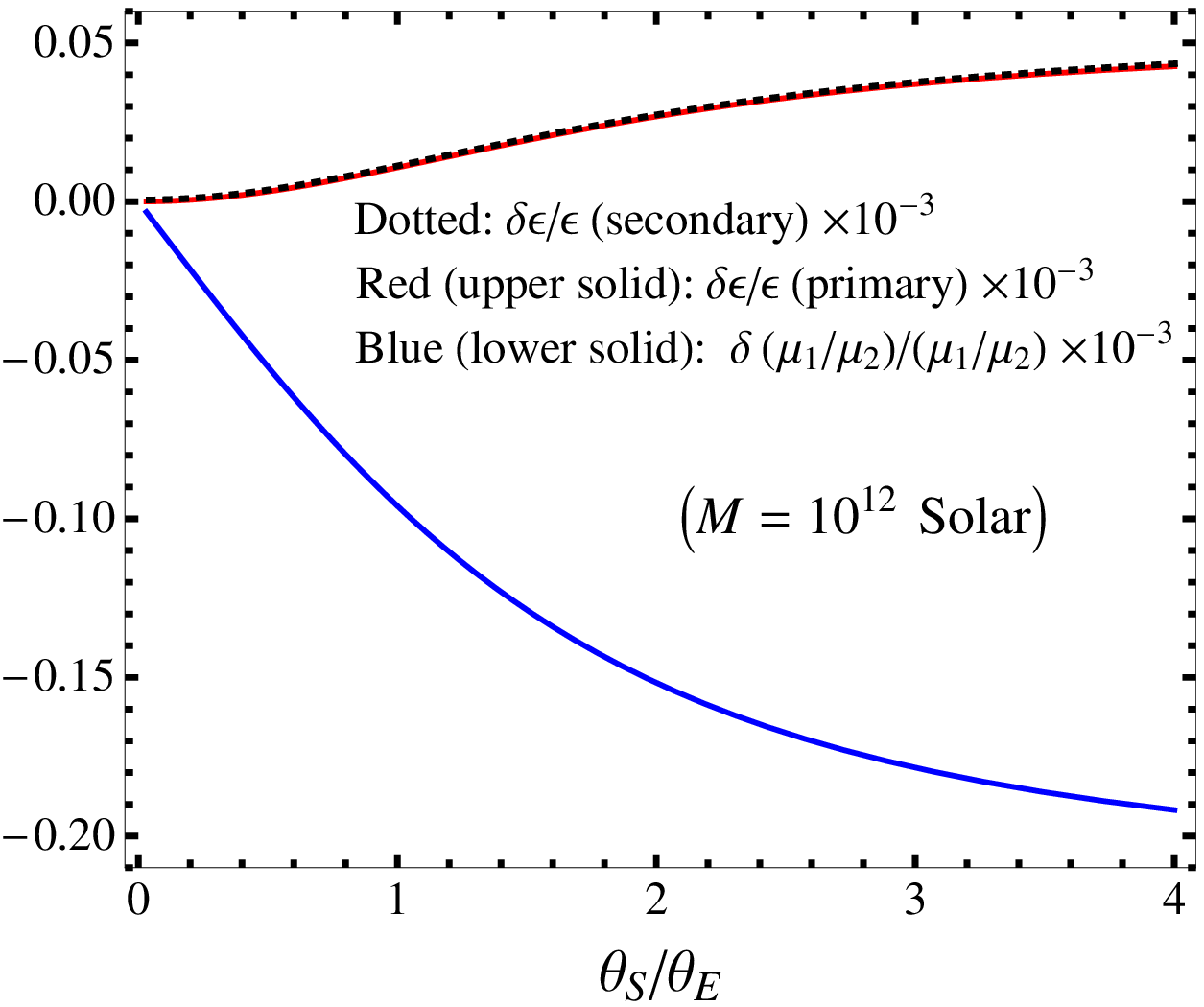}
\end{array}$
\end{center}
\caption{Linear corrections to Schwarzschild lensing caused by the finite range of embedding--- the magnification ratio $\mu_1/\mu_2$ and the ellipticity $\epsilon$ are plotted as a function of source position.  The cosmological parameters and redshifts are same as in Figs. 2 and 3. }
\label{fig:Mag-Ellip}
\end{figure*}

The standard lensing results are obtained by putting $\cos\pht=1$ and $\sin\pht=0$ in the above.
Deviations from standard image amplification $\mu$ and the image ellipticity $\epsilon$ caused by embedding are shown in Fig.\,\ref{fig:Mag-Ellip}.  The left panel is for a deflector mass ${\rm m}=10^{15} \>M_{\odot}$ and the right is for ${\rm m}=10^{12} \>M_{\odot}.$ In each plot, the red solid and the (identical to accuracy shown) black dotted (upper) curves show the corrections in ellipticity, \ie $\delta \epsilon/\epsilon$ for the primary and secondary images. The solid blue (lower) curve is the relative correction in the magnification ratio, \ie $\delta (\mu_1/\mu_2)/(\mu_1/\mu_2).$ For the ${\rm m}=10^{15} \>M_{\odot}$  case, the correction in  ellipticity can be as large as $0.03\%$, and the correction in magnification ratio can be as large as $-0.17\%.$ For the ${\rm m}=10^{12} \>M_{\odot}$  case, the correction in  ellipticity can be as large as $0.004\%$, and the correction in magnification ratio can be as large as $-0.019\%.$

\section{Conclusions}
We have given a lens equation (\ref{lens-eq-1}) valid for use on highly concentrated lenses (point masses) which are embedded into the otherwise spatially homogeneous  and flat background FLRW cosmology. We have also given the additional equations necessary to iteratively solve this embedded lens equation and have outlined a procedure for doing so. As an example we have looked at differences in strong lensing predictions made by this new theory as compared to the conventional theory. We used a large galaxy size lens (${\rm m}=10^{12} \>M_{\odot}$) and  a rich cluster size lens (${\rm m}=10^{15} \>M_{\odot}$) and found, as was suggested before in \cite{Kantowski10, Chen10}, that  predictions for strong lensing effects made by embedded lens theory differs by less than 1\% from predictions made by the conventional theory. In Section II we looked at image angle differences and in Section III we looked at lowest order analytic expressions for image differences. We expect more significant effects to occur for weak lensing applications where impact distances are much larger and where shielding effects ($\cos^3\pht$) are more significant.

Work on this project was partially supported by NSF grant AST-0707704 and US DOE Grant DE-FG02- 07ER41517 and B. Chen wishes to thank the University of Oklahoma Foundation.

\label{lastpage}


\begin{thebibliography}{breitestes Label}


\bibitem[Einstein \& Straus(1945)]{Einstein45} A. Einstein \& E. G. Straus, Rev. Mod. Phys., 17, 120 (1945).

\bibitem[Sch\"ucking(1954)]{Schucking54} E. Sch\"ucking,  Z. Phys., 137, 595 (1954).

\bibitem[Kantowski(1969)]{Kantowski69} R. Kantowski,   \apj, 155, 89 (1969).

\bibitem[Dyer \& Roeder(1974)]{Dyer74} C. C. Dyer \& R. C. Roeder,  \apj, 189, 167 (1974).

\bibitem[Sch\"{u}cker(2009a)]{Schucker09a} T. Sch\"ucker, Gen. Relativ. Gravit., 41, 67 (2009).

\bibitem[Kottler(1918)]{Kottler18} F. Kottler, Ann. Phys. (Leipzig), 361, 401 (1918).

\bibitem[Kantowski \etal (2010)]{Kantowski10} R. Kantowski, B. Chen  \& X. Dai,  \apj, 718, 913 (2010).

\bibitem[Chen \etal (2010)]{Chen10} B. Chen, R. Kantowski  \& X.  Dai,  \prd, 82, 043005 (2010).

\bibitem[Rindler \& Ishak(2007)]{Rindler07} W. Rindler \& M. Ishak, \prd, 76, 043006 (2007).

\bibitem[Sch\"{u}cker(2009b)]{Schucker09b} T. Sch\"ucker,  Gen. Relativ. Gravit., 41, 1595 (2009).

\bibitem[Sch\"{u}cker(2010)]{Schucker10}  T. Sch\"ucker, arXiv:1006.3234 (2010).

\bibitem[Boudjemaa et al.(2011)]{Boudjemaa} K.-E. Boudjemaa, M. Guenouche  \& S. R. Zouzou,  Gen. Relativ. Gravit., 43, 1707 (2011).

\bibitem[Ishak et al.(2010)]{Ishak10a} M. Ishak, W. Rindler \& J. Dossett, Mon. Not. R. Astron. Soc., 403, 21521 (2010).

\bibitem[Ishak \& Rindler(2010)]{Ishak10b} M. Ishak \& W. Rindler,   Gen. Relativ. Gravit., 42, 2247 (2010).

\bibitem[Schneider \etal (1992)]{Ehlers} P. Schneider, J Ehlers \& E. E.  Falco,  {\it Gravitational Lenses} (Springer-Verlag, Berlin, 1992).

\bibitem[Bourassa \& Kantowski(1975)]{Bourassa75} R. R. Bourassa \& R. Kantowski,  \apj, 195, 13 (1975).

\bibitem[Ishak(2008)]{Ishak08a} M. Ishak,  \prd, 78, 103006 (2008).

\bibitem[Sereno (2009)]{Sereno09}  M. Sereno,  \prl, 102, 021301 (2009).

\bibitem[Sereno (2008)]{Sereno08} M. Sereno,  \prd, 77, 043004 (2008).

\bibitem[Ishak et al.(2008)]{Ishak08b} M. Ishak, W. Rindler, J. Dossett, J. Moldenhauer \& C. Allison,  Mon. Not. R. Astron. Soc., 388, 1279 (2008).


\end{thebibliography}
\end{document}